\documentclass{article}
\usepackage{amsmath}
\usepackage{amsfonts}
\usepackage[ignoreall]{geometry}

\newtheorem{theorem}{Theorem}

\newtheorem{definition}[theorem]{Definition}

\newtheorem{proposition}[theorem]{Proposition}

\newenvironment{proof}[1][Proof]{\noindent\textbf{#1.} }{\ \rule{0.5em}{0.5em}}
\input{tcilatex}
\begin{document}

\title{On the closure property of Lepage equivalents of Lagrangians}
\author{Nicoleta VOICU, Stefan GAROIU, Bianca VASIAN \\
Transilvania University of Brasov, Romania}
\date{}
\maketitle

\begin{abstract}
Lepage equivalents of Lagrangians are a higher order, field-theoretical
generalization of the notion of Poincar\'{e}-Cartan form from mechanics and
play a similar role: they give rise to a geometric formulation (and to a
geometric understanding) of the calculus of variations.

A long-standing open problem is the determination, for field-theoretical
Lagrangians $\lambda $ of order greater than one, of a Lepage equivalent $%
\Phi _{\lambda }$ with the so-called \textit{closure property:\ }$\Phi
_{\lambda }$ is a closed differential form if and only if $\lambda $ has
vanishing Euler-Lagrange expressions.

The present paper proposes a solution to this problem, for general
Lagrangians of order $r\geq 1.$ The construction is a local one; yet, we
show that in most of the cases of interest for physical applications, the
obtained Lepage equivalent $\Phi _{\lambda }$ is actually globally defined.

A variant of this construction, which is convenient in the cases when $%
\lambda $ is a reducible Lagrangian, is also introduced. In particular, for
reducible Lagrangians of order two, the obtained Lepage equivalents are of
order one.
\end{abstract}

\textbf{Corresponding author's e-mail: }nico.voicu@unitbv.ro



\section{Introduction}

In classical mechanics, the Poincar\'{e}-Cartan form%
\begin{equation}
\Theta _{L}=Ldt+\dfrac{\partial L}{\partial \dot{q}^{\sigma }}(dq^{\sigma }-%
\dot{q}^{\sigma }dt)  \label{classical Cartan form}
\end{equation}%
associated to a Lagrangian $L=L(t,q^{\sigma },\dot{q}^{\sigma })$ is a
fundamental concept, giving rise to a geometric understanding of the
calculus of variations. Similarly, its higher-order, field theoretical
analogue, mainly known under the name of \textit{Lepage equivalent }of a
Lagrangian \cite{Krupka1}, \cite{KKS}, \cite{Giachetta}, allows for a
concise and elegant description of the apparatus of variational calculus,
solely in terms of operations with differential forms. In view of such a
description, the Lagrangian itself will be regarded as a differential form $%
\lambda $ on a certain jet bundle, rather than as a function (for instance,
in the above example, $\lambda =Ldt$).

But, whereas in mechanics, the Poincar\'{e}-Cartan form is unique, in field
theory, any given Lagrangian $\lambda $ admits multiple Lepage equivalents $%
\theta _{\lambda }$, exhibiting different features. One of the most
desirable such features is the so-called \textit{closure property:\ }

\begin{center}
$\lambda $ is variationally trivial $~\Leftrightarrow ~\ \ d\theta _{\lambda
}=0$.
\end{center}

Once the closure property is satisfied, all Lagrangians producing the same
Euler-Lagrange equations will be characterized by one and the same $d\theta
_{\lambda },$ in other words, $d\theta _{\lambda }$ will capture a specific 
\textit{dynamics}, not just a particular Lagrangian\footnote{%
Moreover, it can be reasonably argued, \cite{Grigore-Popp}, \cite{KKS},
that, in a lot of considerations (especially, having in view a Hamiltonian
picture), it is the exterior derivative $d\theta _{\lambda },$ rather than $%
\theta _{\lambda }$ itself, that plays the key role.}.

This property was initially motivated by the study of symmetries of the
Euler-Lagrange form, see \cite{Betounes}, \cite{Betounes-symmetry}, but it
is a very promising one in at least two other directions:

- Geometric formulation of Hamiltonian field theory: given a Lagrangian form 
$\lambda $, a Hamiltonian form $H_{\lambda }$ is constructed via the
exterior derivative $d\theta _{\lambda }$ - and generally, it is not
guaranteed that Lagrangians that produce the same Euler-Lagrange equations
will also lead to the same Hamilton equations. This drawback is eliminated
if the mapping $\lambda \mapsto \theta _{\lambda }$ is $\mathbb{R}$-linear
and satisfies the closure property.

- Variational sequences (e.g., \cite{Krupka-book}), where it offers an
elegant characterization of the kernel of the Euler-Lagrange mapping.

\bigskip

The closure property is notoriously obeyed in mechanics by the Poincar\'{e}%
-Cartan form $\Theta _{L}$, both in the first order case (\ref{classical
Cartan form}) and for higher order Lagrangians. But, in \textit{field theory}%
, finding Lepage equivalents with the closure property has been for many
years an open problem. Actually, to the best of our knowledge, mappings $%
\lambda \mapsto \theta _{\lambda }$ obeying it are only known in some very
specific situations:

1. First order Lagrangians. In this case, a globally defined Lepage
equivalent with the desired feature, called the \textit{fundamental Lepage
equivalent }$\rho _{\lambda }$\textit{,} was introduced by Krupka, \cite%
{Krupka-fundamental} and rediscovered by Bethounes, \cite{Betounes}; for
first order homogeneous Lagrangians, a similar notion was introduced by
Urban and Brajercik, \cite{Urban}.

2. Homogeneous Lagrangians with two independent variables; in this case, an
extension of the fundamental form $\rho _{\lambda }$ was constructed by
Saunders and Crampin, \cite{Saunders}.

\bigskip

In the present paper, we propose a general procedure which solves this
problem (at least, locally) for general Lagrangians $\lambda ,$ of any order 
$r\geq 1$. Our construction relies on a different idea than the first order
construction in \cite{Krupka-fundamental}, as it uses as a raw material,
another Lepage equivalent, called the \textit{principal Lepage equivalent}, 
\cite{Krupka1} - which is much simpler; more specifically, it is 1-contact,
whereas the fundamental Lepage equivalent $\rho _{\lambda }$ has a higher
degree of contactness.

The principal Lepage equivalent, as it stands, does not obey the closure
property, but we show that it can be tailored in such a way as to eliminate
this drawback, as follows. To any Lagrangian $\lambda $ over a given fibered
chart domain, one can canonically attach, on the respective chart domain, a
dynamically equivalent\textit{\ }Lagrangian:\ the so-called \textit{%
Vainberg-Tonti Lagrangian} $\lambda _{VT}$ of the Euler-Lagrange form of $%
\lambda $. The difference between $\lambda $ and $\lambda _{VT}$ is thus a
trivial Lagrangian, which can be written, \cite{Krupka-book}, up to pullback
by the corresponding jet projections, as 
\begin{equation}
\lambda =\lambda _{VT}+hd\alpha ,  \label{canonical_splitting_Intro}
\end{equation}%
where $h$ denotes the horizontalization operator and $d\alpha $ is uniquely
determined, via a specific homotopy operator.

Using the above decomposition, we define 
\begin{equation}
\Phi _{\lambda }:=\Theta _{\lambda _{VT}}+d\alpha ,
\label{canonical_Lepage_Intro}
\end{equation}%
where $\Theta _{\lambda _{VT}}$ is the principal Lepage equivalent of $%
\lambda _{VT}$ and the equality should be, again, understood up to pullback
by some jet projections. The differential form $\Phi _{\lambda }$ is a
Lepage equivalent of $\lambda ,$ which we call \textit{canonical. }As for
variationally trivial Lagrangians $\lambda ,$ the associated Vainberg-Tonti
Lagrangian $\lambda _{VT}$ identically vanishes, the above recipe guarantees
that, in this case, $\Phi _{\lambda }=d\alpha ,$ hence it is closed.
Moreover, $\Phi _{\lambda }$ is 1-contact.

As, typically, both the Vainberg-Tonti Lagrangian and the principal Lepage
equivalent are only defined over a coordinate chart, a natural question is
whether (or, rather, when) is $\Phi _{\lambda }$ globally defined. A
complete answer to this question is still to be investigated; yet, we show
that, for globally defined Lagrangians of order $r\leq 2$ on tensor bundles,
having second order Euler-Lagrange equations - which represent most of the
cases of interest for physical theories - $\Phi _{\lambda }$ is actually
globally well defined.

A variant of the above construction, which is convenient in the case when $%
\lambda $ is locally equivalent to a lower order Lagrangian $\lambda
^{\prime }$, is to consider in (\ref{canonical_Lepage_Intro}), instead of
the Vainberg-Tonti Lagrangian $\lambda _{VT}$, a reduced Lagrangian\textit{\ 
}$\lambda ^{\prime }.$ This leads to (generally, non-unique) Lepage
equivalents $\phi _{\lambda }$, which we will call \textit{reduced}; and, if
we can ensure that $\lambda ^{\prime }$ is truly of minimal order, the
obtained reduced Lepage equivalent will still possess the closure property.
In particular, for reducible second order Lagrangians, any reduced Lepage
equivalent will be of order 1.

The article is structured as follows. Section 2 is a somewhat didactic
presentation of the known results and notions to be used in the following.
In Sections 3 and 4, we introduce the canonical, respectively, minimal
Lepage equivalents. Section 5 is devoted to examples and Section 6, to
conclusions and future work perspectives.

\section{Preliminaries}

In the following, we present the technical ingredients to be used in our
construction. The notions and results presented in this section can be found
in more detail, e.g., in the book by Krupka, \cite{Krupka-book}.

In the approach we present below, a Lagrangian is regarded as a differential
form on a certain jet bundle of a fibered manifold $(Y,\pi ,X);$ in physics,
the total space $Y$ of the respective fibered manifold is interpreted as the
configuration space\textit{\ }of a given system, the base manifold $X$ is
typically interpreted as space (or spacetime) and sections of $\pi $ are
interpreted as fields. The case $\dim X=1$ corresponds to mechanics and, in
this case, sections of $Y$ are curves.

\subsection{Lagrangians, Lepage equivalents and first variation formula}

A fibered manifold is a triple $\left( Y,\pi ,X\right) ,$ where $X$, $Y$ are
smooth manifolds, with $\dim X=n,$ $\dim Y=m+n,$ and $\pi :Y\rightarrow X$
is a surjective submersion; the base manifold $X$ is assumed to be
orientable. On a fibered manifold, there exists an atlas consisting of
fibered charts $(V,\psi )$, $\psi =(x^{i},y^{\sigma })$, such that $\pi $ is
represented in coordinates as $\pi :\left( x^{i},y^{\sigma }\right) \mapsto
\left( x^{i}\right) .$ We will denote by $\Gamma (Y)$ the set of local
sections $\gamma :U\rightarrow Y$ (where $U\subset X$ is open); elements $%
\gamma \in \Gamma (Y)$ are represented in a fibered chart as $\gamma :\left(
x^{i}\right) \mapsto (x^{i},y^{\sigma }\left( x^{i}\right) ).$

Each fibered chart $(V,\psi )$ induces a fibered chart $(V^{r},\psi ^{r}),$ $%
\psi ^{r}=(x^{i},y^{\sigma },y_{~j_{1}}^{\sigma
},...,y_{~j_{1}j_{2}...j_{r}}^{\sigma })$ on the $r$-jet prolongation $%
J^{r}Y.$ In these charts, the canonical projections $\pi
^{r,s}:J^{r}Y\rightarrow J^{s}Y,\ J_{x}^{r}\gamma \mapsto J_{x}^{s}\gamma $ (%
$r>s,$ $J^{0}Y:=Y$) are represented as: $(x^{i},y^{\sigma
},y_{~j_{1}}^{\sigma },...,y_{~j_{1}j_{2}...j_{r}}^{\sigma })\mapsto
(x^{i},y^{\sigma },y_{~j_{1}}^{\sigma },...,y_{~j_{1}j_{2}...j_{s}}^{\sigma
})$ and the projection $\pi ^{r}:J^{r}Y\rightarrow X,J_{x}^{r}\gamma \mapsto
x,~$as: $(x^{i},y^{\sigma },y_{~j_{1}}^{\sigma
},...,y_{~j_{1}j_{2}...j_{r}}^{\sigma })\mapsto (x^{i}).$ By $\Omega
_{k}(W^{r})$ and $\Omega (W^{r})$ we will denote the set of $k$-forms
(respectively, of all differential forms) defined on $W^{r}:=J^{r}W\subset
J^{r}Y,$ where $W\subset Y$ is open.

Also, for the simplicity of writing, if there is no risk of confusion, we
will sometimes identify forms $\rho $ with their pullbacks $\left( \pi
^{s,r}\right) ^{\ast }\rho ,$ $s\geq r;$ that is, instead of $\left( \pi
^{s,r}\right) ^{\ast }\rho =\theta ,$ we may simply write $\rho =\theta .$

\bigskip

\textbf{Horizontal forms and contact forms on }$J^{r}Y$\textbf{. }A
differential form $\rho \in \Omega _{k}(W^{r})$ is called $\pi ^{r}$%
-horizontal, if $\rho (\Xi _{1},...,\Xi _{k})=0$ whenever one of the vector
fields $\Xi _{i},$ $i=\overline{1,k},$ is $\pi ^{r}$-vertical (i.e., $T\pi
^{r}(\Xi _{i})=0$); in a fibered chart, any horizontal form on $\Omega
_{k}(W^{r})$ is expressed as a linear combination of the wedge products $%
dx^{i_{1}}\wedge dx^{i_{2}}\wedge ...\wedge dx^{i_{k}}$. Yet, for a more
compact writing, it is advantageous to use the following locally defined
forms:%
\begin{eqnarray}
&&\omega _{0}:=dx^{1}\wedge ...\wedge dx^{n},~\ \ \ \omega _{i}:=\mathbf{i}%
_{\partial _{i}}\omega _{0}=\left( -1\right) ^{i-1}dx^{1}\wedge ...\wedge 
\widehat{dx^{i}}\wedge ...\wedge dx^{n},  \label{omega_0} \\
&&\omega _{i_{1}...i_{k}}:=\mathbf{i}_{\partial _{i_{k}}}\mathbf{i}%
_{\partial _{i_{k-1}}}...\mathbf{i}_{\partial _{i_{1}}}\omega _{0},
\label{omega_i1_ik}
\end{eqnarray}%
where $\mathbf{i}$ denotes interior product. This way, any $\pi ^{r}$%
-horizontal $k$-form on $W^{r}$ ($k\leq n$) will have a coordinate
expression:%
\begin{equation}
\rho =\dfrac{1}{k!}A^{i_{1}...i_{k}}\omega _{i_{1}...i_{k}},
\label{horizontal_form}
\end{equation}%
where $A^{i_{1}...i_{k}}$ are smooth functions of $x^{i},y^{\sigma
},y_{~j_{1}}^{\sigma },...,y_{~j_{1}j_{2}...j_{r}}^{\sigma }$. The set of $%
\pi ^{r}$-horizontal $k$-forms over $W$ will be denoted by $\Omega
_{k,X}(W^{r}).$

Similarly, one can define $\pi ^{r,s}$\textit{-horizontal} forms, $0\leq
s\leq r$; locally, these are generated by wedge products of $%
dx^{i},dy^{\sigma },...,dy_{~j_{1}...j_{s}}^{\sigma }.$

\bigskip

A form $\theta \in \Omega (W^{r})$ is called a \textit{contact form }if it
vanishes along all prolonged sections, i.e., $J^{r}\gamma ^{\ast }\theta =0,$
$\forall \gamma \in \Gamma (Y).$ For instance: 
\begin{eqnarray}
\omega ^{\sigma } &=&dy^{\sigma }-y_{~j}^{\sigma }dx^{j},
\label{contact basis} \\
\omega _{~i_{1}i_{2}...i_{k}}^{\sigma } &=&dy_{~i_{1}i_{2}...i_{k}}^{\sigma
}-y_{~i_{1}i_{2}...i_{k}j}^{\sigma }dx^{j},  \label{contact_basis_1}
\end{eqnarray}%
where $k\leq r-1,$ represent contact forms; more precisely, they are
elements of a local basis of $\Omega _{1}(W^{r}),$ called the \textit{%
contact basis: }$\{dx^{i},\omega ^{\sigma },....,\omega
_{~i_{1}...i_{r-1}}^{\sigma },dy_{~i_{1}...i_{r}}^{\sigma }\}.$

A $k$-form $\theta \in \Omega _{k}(W^{r})$ is called $l$\textit{-contact (}$%
l\leq k$)\textit{\ }if $\mathbf{i}_{\Xi _{1}}\mathbf{i}_{\Xi _{2}}...\mathbf{%
i}_{\Xi _{l}}\theta $ is horizontal whenever the vector fields $\Xi
_{1},...,\Xi _{l}$ are all $\pi ^{r}$-vertical; in the contact basis, in the
expression of an $l$-contact $k$-form, each term contains precisely $l$ of
the contact 1-forms (\ref{contact basis})-(\ref{contact_basis_1}) (and $k-l$
of the forms $dx^{i}$).

Rising to $J^{r+1}Y,$ any $\rho \in \Omega _{k}(W^{r})$ can be uniquely
split into a $\pi ^{r}$-horizontal part $h\rho \in \Omega _{k}(W^{r+1})$ and
a contact part $p\rho \in \Omega _{k}(W^{r+1})$:%
\begin{equation}
(\pi ^{r+1,r})^{\ast }\rho =h\rho +p\rho ;  \label{first_canonical_decomp}
\end{equation}%
the contact part can be in its turn decomposed as%
\begin{equation*}
p\rho =p_{1}\rho +...+p_{k}\rho ,
\end{equation*}%
where the form $p_{l}\rho $ is $l$-contact, $l=1,...,k$. On the other hand,
the mapping $\Omega (W^{r})\rightarrow \Omega (W^{r+1}),\rho \mapsto h\rho $
in (\ref{first_canonical_decomp}), is a\ morphism of exterior algebras,
called \textit{horizontalization; }it acts on functions $f:J^{r}Y\rightarrow 
\mathbb{R}$ as:%
\begin{equation*}
hf=f\circ \pi ^{r+1,r}~,~\ hdf=d_{i}fdx^{i},
\end{equation*}%
where $d_{i}$ denotes total $x^{i}$-derivative (of order $r+1$): $%
d_{i}f:=\partial _{i}f+\dfrac{\partial f}{\partial y^{\sigma }}%
y_{~i}^{\sigma }+...\dfrac{\partial f}{\partial y_{~j_{1}...j_{r}}^{\sigma }}%
y_{~j_{1}...j_{r}i}^{\sigma }.$ For instance, $hdx^{i}=dx^{i},~\ hdy^{\sigma
}=y_{~i}^{\sigma }dx^{i},...,hdy_{~j_{1}...j_{r}}^{\sigma
}=y_{~j_{1}...j_{r}i}^{\sigma }dx^{i}.$

The horizontal part $h\rho $ is the only one that survives of $\rho $ when
pulled back by prolonged sections $J^{r}\gamma $; more precisely, for any $%
\gamma \in \Gamma (Y):$\ $J^{r}\gamma ^{\ast }\rho =J^{r+1}\gamma ^{\ast
}(h\rho ).$

\bigskip

Here are two important classes of differential forms on $W^{r}\subset
J^{r}Y: $

\begin{enumerate}
\item \textit{Lagrangians}: A Lagrangian of order $r$ is, by definition, a $%
\pi ^{r}$-horizontal form $\lambda \in \Omega _{n,X}(W^{r})$ of rank $n=\dim
X;$\textit{\ }in fibered coordinates, a Lagrangian is expressed as:%
\begin{equation}
\lambda =\mathcal{L}\omega _{0},~\ \ \ \ \ \ \mathcal{L=L}(x^{i},y^{\sigma
},...,y_{i_{1}...i_{r}}^{\sigma }),  \label{Lagrangian}
\end{equation}%
where $\omega _{0}$ is as in (\ref{omega_0}).

\item \textit{Source forms }(or \textit{dynamical forms, \cite{Rossi}}) are
defined as $\pi ^{r,0}$-horizontal, 1-contact $\left( n+1\right) $-forms; in
fibered coordinates, any source form $\varepsilon \in \Omega _{n+1}(W^{r})$
is represented as:%
\begin{equation*}
\varepsilon =\varepsilon _{\sigma }\omega ^{\sigma }\wedge \omega _{0},~\ \
\ \ \ \varepsilon _{\sigma }=\varepsilon _{\sigma }(x^{i},y^{\sigma
},...,y_{i_{1}...i_{r}}^{\sigma }),
\end{equation*}%
where $\omega ^{\sigma }$ are as in (\ref{contact basis}). The most
prominent particular case of source forms are Euler-Lagrange forms of
Lagrangians, to be discussed below.
\end{enumerate}

\bigskip

\textbf{Lepage equivalents and first variation formula. }Consider a
Lagrangian $\lambda \in \Omega _{n,X}(W^{r})$ of order $r.$ The action%
\textit{\ }attached to $\lambda $ and to a compact domain $D\subset X\ $is
the function $S:\Gamma (Y)\rightarrow \mathbb{R},$ given by: 
\begin{equation}
S(\gamma )=\underset{D}{\int }J^{r}\gamma ^{\ast }\lambda .  \label{action}
\end{equation}%
The variation of $S$ under the flow of a $\pi $-projectable vector field $%
\Xi =\xi ^{i}\partial _{i}+\Xi ^{\sigma }\partial _{\sigma }$ on $Y$ is then:%
\begin{equation}
\delta _{\Xi }S(\gamma )=\underset{D}{\int }J^{r}\gamma ^{\ast }\mathfrak{L}%
_{J^{r}\Xi }\lambda ,  \label{variation}
\end{equation}%
where the symbol $\mathfrak{L}$ stands for Lie derivative.

\bigskip

A\textit{\ Lepage equivalent}\ \textit{of a Lagrangian}\footnote{%
A related, but distinct, notion is the one of Lepage equivalent\textit{\ of
a source form} $\varepsilon $\textit{, }\cite{Rossi}. In the following,
whenever we refer to Lepage equivalents, we will automatically mean Lepage
equivalents of \textit{Lagrangians}.} $\lambda \in \Omega _{n,X}(W^{r})$ is
an $n$-form $\theta _{\lambda }\in \Omega _{n}(W^{s})$ on some jet
prolongation $J^{s}Y,$ with the following properties:

(1) $\theta _{\lambda }$ and $\lambda $ \textit{define the same variational
problem}, i.e., up to the corresponding jet projections: 
\begin{equation}
h\theta _{\lambda }=\lambda .  \label{Lepage_def1}
\end{equation}

(2)\ $\mathcal{E}_{\lambda }:=p_{1}d\theta _{\lambda }$ is a source form.

\bigskip

Condition (1)\ should be understood as follows. For any section $\gamma \in
\Gamma (Y),$ there holds: $J^{s}\gamma ^{\ast }\theta _{\lambda
}=J^{r}\gamma ^{\ast }\lambda $, which means that we can substitute $%
J^{s}\gamma ^{\ast }\theta _{\lambda }$ for $J^{r}\gamma ^{\ast }\lambda $
into the action (\ref{action}). Condition (2) actually states that $%
p_{1}d\theta _{\lambda }$ must be locally generated by $\omega ^{\sigma }$
and $dx^{i}$ alone (no higher order elements $\omega
_{~i_{1}...i_{k}}^{\sigma }$ of the contact basis).

\bigskip

Given any Lepage equivalent $\theta _{\lambda }\in \Omega _{n}(W^{s})$ of $%
\lambda ,$ Cartan's formula $\mathfrak{L}_{J^{r}\Xi }\theta _{\lambda }=d%
\mathbf{i}_{J^{r}\Xi }\theta _{\lambda }+\mathbf{i}_{J^{r}\Xi }d\theta
_{\lambda }$ applied to the integrand in (\ref{variation}) yields: 
\begin{equation}
J^{r}\gamma ^{\ast }(\mathfrak{L}_{J^{r}\Xi }\lambda )=J^{s+1}\gamma ^{\ast }%
\mathbf{i}_{J^{s+1}\Xi }\mathcal{E}_{\lambda }+d(J^{s}\gamma ^{\ast }%
\mathcal{J}^{\Xi }),  \label{first_variation_refined}
\end{equation}%
where:

\begin{enumerate}
\item[\textit{(i)}] the source form $\mathcal{E}_{\lambda }=p_{1}d\theta
_{\lambda }\in \Omega _{n}(W^{s+1})$ (of order $s+1\leq 2r$) is the \textit{%
Euler-Lagrange form} of $\lambda $, locally given by\textit{\ }%
\begin{eqnarray}
\mathcal{E}_{\lambda } &=&\mathcal{E}_{\sigma }(\lambda )\omega ^{\sigma
}\wedge \omega _{0},  \notag \\
\mathcal{E}_{\sigma }(\lambda ) &=&\dfrac{\delta \mathcal{L}}{\delta
y^{\sigma }}=\dfrac{\partial \mathcal{L}}{\partial y^{\sigma }}-d_{i}\dfrac{%
\partial \mathcal{L}}{\partial y_{~i}^{\sigma }}%
+...+(-1)^{r}d_{i_{1}}...d_{i_{r}}\dfrac{\partial \mathcal{L}}{\partial
y_{~i_{1}...i_{r}}^{\sigma }}.  \label{Euler-Lagrange_expr}
\end{eqnarray}%
The Euler-Lagrange form $\mathcal{E}_{\lambda }$ does not depend on the
choice of the Lepage equivalent $\theta _{\lambda }$.

A local section $\gamma \in \Gamma (Y)$ is critical for the Lagrangian $%
\lambda $ if and only if, for any $\pi $-vertical vector field $\Xi \in 
\mathcal{X}(Y),$ there holds:%
\begin{equation}
J^{s+1}\gamma ^{\ast }(\mathbf{i}_{J^{s+1}\Xi }\mathcal{E}_{\lambda })=0,
\label{coord-free-EL-eq}
\end{equation}%
which in coordinates, becomes equivalent to the relations $\mathcal{E}%
_{\sigma }(\lambda )\circ J^{s+1}\gamma =0.$

\item[\textit{(ii)}] $\mathcal{J}^{\Xi }:=h\mathbf{i}_{J^{s}\Xi }\theta
_{\lambda }\in \Omega _{n-1}(J^{s+1}Y)$ is interpreted as a Noether current.
\end{enumerate}

\bigskip

Every Lagrangian $\lambda \in \Omega _{n,X}(W^{r})$ admits Lepage
equivalents. The most frequently used one, called the \textit{principal
Lepage equivalent, }is a 1-contact form of order $\leq 2r-1$ and it is given
by: 
\begin{equation}
\Theta _{\lambda }=\mathcal{L}\omega _{0}+\overset{r-1}{\underset{k=0}{\sum }%
}(\overset{r-1-k}{\underset{l=0}{\sum }}(-1)^{l}d_{p_{1}}...d_{p_{l}}\dfrac{%
\partial \mathcal{L}}{\partial y_{~j_{1}...j_{k}p_{1}...p_{l}i}^{\sigma }}~\
\ \omega _{j_{1}...j_{k}}^{\sigma })\wedge \omega _{i}.
\label{Poincare-Cartan_general}
\end{equation}%
Generally, the principal Lepage equivalent $\Theta _{\lambda }$ is defined
only locally; yet, for first and second order Lagrangians, it is globally
defined whenever $\lambda $ itself is globally defined, \cite{Krupka-book}, 
\cite{Rossi}\textit{. }

In particular, for second order Lagrangians $\lambda =\mathcal{L}%
(x^{i},y^{\sigma },y_{~i}^{\sigma },y_{~ij}^{\sigma })\omega _{0},$ the
above formula gives:%
\begin{equation}
\Theta _{\lambda }=\mathcal{L}\omega _{0}+B_{~\sigma }^{i}\omega ^{\sigma
}\wedge \omega _{i}+B_{~\sigma }^{ij}\omega _{~j}^{\sigma }\wedge \omega
_{i},  \label{2nd_order_Theta}
\end{equation}%
where:%
\begin{equation}
B_{~\sigma }^{i}=\dfrac{\partial \mathcal{L}}{\partial y_{~i}^{\sigma }}%
-d_{j}(\dfrac{\partial \mathcal{L}}{\partial y_{~ij}^{\sigma }}),~\ \ \
B_{~\sigma }^{ij}=\dfrac{\partial \mathcal{L}}{\partial y_{~ij}^{\sigma }}.
\label{B_general}
\end{equation}

\begin{proposition}
\label{remark_order_Theta}If $\lambda =\mathcal{L}\omega _{0}\in \Omega
_{n,X}(W^{r})$ is \textit{affine} in the highest order variables $%
y_{~i_{1}...i_{r}}^{\sigma },$ the order of $\Theta _{\lambda }$ is
actually, at most $2r-2$.
\end{proposition}

\begin{proof}
The statement follows immediately by inspecting the highest order term $%
d_{p_{1}}...d_{p_{r-1}}\dfrac{\partial \mathcal{L}}{\partial
y_{~p_{1}...p_{r-1}i}^{\sigma }}$ of $\Theta _{\lambda }.$
\end{proof}

Any other Lepage equivalent $\theta _{\lambda }$ of $\lambda $ can be
locally decomposed as:%
\begin{equation}
\theta _{\lambda }=\Theta _{\lambda }+d\nu +\mu ,
\label{general_Lepage_equiv}
\end{equation}%
where $\nu $ is 1-contact and $\mu $ is 2-contact. In particular, any 
\textit{1-contact }Lepage equivalent of $\lambda $\textit{\ }can be
expressed, \cite{Rossi}, up to the corresponding jet projections, as:%
\begin{equation}
\theta _{\lambda }=\Theta _{\lambda }+p_{1}d\nu .
\label{1-contact_Lepage_equivs}
\end{equation}

\subsection{The closure property\label{Section: closure}}

A Lagrangian $\lambda \in \Omega _{n,X}(W^{r})$ is called \textit{trivial}
(or \textit{null}) if its Euler-Lagrange form $\mathcal{E}_{\lambda }$
vanishes identically. It is known, e.g., \cite{Krupka-book}, p. 123, that $%
\lambda $ is trivial if and only if, for any fibered chart domain $%
V^{r}=J^{r}V\subset W^{r},$ there exists an $\left( n-1\right) $-form $%
\alpha \in \Omega _{n-1}(V^{r-1})$ of order $r-1,$ such that:

\begin{equation}
\lambda =hd\alpha .  \label{trivial Lagrangian}
\end{equation}

\bigskip

A mapping $\theta :\Omega _{n,X}(W^{r})\rightarrow \Omega _{n}(W^{s})$
attaching to any Lagrangian $\lambda \in \Omega _{n,X}(J^{r}Y),$ a Lepage
equivalent $\theta _{\lambda }$ of some order $s,$ is said, \textit{\cite%
{Saunders},} to have the \textit{closure property, }if: 
\begin{equation}
\lambda \text{ \textit{- trivial} }\Rightarrow ~d\theta _{\lambda }=0,
\label{closure property}
\end{equation}%
for all $\theta _{\lambda }$ in the image of $\lambda .$

\bigskip

\textbf{Remark. }The converse implication:\ $d\theta _{\lambda }=0$ $%
\Rightarrow \lambda $ \textit{- trivial}, is true for \textit{any }Lepage
equivalent $\theta _{\lambda },$ since $d\theta _{\lambda }=0$ implies $%
\mathcal{E}_{\lambda }=p_{1}d\theta _{\lambda }=0$; hence, whenever it
holds, (\ref{closure property}) is actually an equivalence.

\bigskip

A first consequence of the closure property is the following.

\begin{proposition}
\label{Prop:d_theta}: If the mapping $\lambda \mapsto \theta _{\lambda
}:\Omega _{n,X}(W^{r})\rightarrow \Omega _{n}(W^{s})$ is $\mathbb{R}$-linear
and has the closure property, then, for any two dynamically equivalent
Lagrangians $\lambda ,\lambda ^{\prime }\in \Omega _{n,X}(W^{r})$:%
\begin{equation}
d\theta _{\lambda }=d\theta _{\lambda ^{\prime }}.  \label{equality_d_theta}
\end{equation}
\end{proposition}

\begin{proof}
Assuming that the Lagrangians $\lambda ,\lambda ^{\prime }\in \Omega
_{n,X}(W^{r})$ are equivalent, it follows that the difference $\lambda
-\lambda ^{\prime }$ is a trivial Lagrangian, hence $d\theta _{\lambda
-\lambda ^{\prime }}=0,$ which, by linearity, implies (\ref{equality_d_theta}%
).
\end{proof}

\bigskip

The closure property is a very convenient one for physical applications, as,
basically, equality (\ref{equality_d_theta}) says that all Lagrangians
describing the same physics will produce the same $d\theta _{\lambda }$;
this can be used, for instance, in:

\begin{enumerate}
\item \textit{Geometric formulation of Hamiltonian field theory} based on
Lepage equivalents,\textit{\ }see, e.g.,\textit{\ }\cite{Krupka-Hamiltonian}%
, \cite{Krupkova-Smetanova}, \cite{Krupkova-2001-Hamiltonian}\textit{.} Fix,
for the moment, an arbitrary Lepage equivalent $\theta _{\lambda }\in \Omega
_{n}(W^{s})$ of a given Lagrangian $\lambda \in \Omega _{n,X}(W^{r}).$ A
local section $\delta $ of the fibered manifold $\left( J^{s}Y,\pi
^{s},X\right) $ is called a \textit{Hamilton extremal }of $\theta _{\lambda
},$ if, for any $\pi ^{s}$-vertical vector field $\xi $ on $J^{s}Y:$%
\begin{equation}
\delta ^{\ast }\mathbf{i}_{\xi }d\theta _{\lambda }=0.
\label{coord-free-Hamilton-eq}
\end{equation}%
Thus, the Hamilton equation (\ref{coord-free-Hamilton-eq}) depends not only
on $\lambda ,$ but also on the choice of the Lepage equivalent $\theta
_{\lambda }.$ In particular, it is not guaranteed that Lagrangians producing
the same Euler-Lagrange equation (\ref{coord-free-EL-eq}) would generally
also produce the same Hamilton equation - which is a major drawback. This
drawback can, yet, be eliminated if the ($\mathbb{R}$-linear) mapping $%
\lambda \mapsto \theta _{\lambda }$ has the closure property, as, in this
case, (\ref{equality_d_theta}) ensures that, for all equivalent Lagrangians $%
\lambda ,\lambda ^{\prime }$, the resulting Hamilton equation (\ref%
{coord-free-Hamilton-eq}) will be the same\footnote{%
Of course, the closure property alone does not guarantee that the resulting
Hamilton equation, albeit unique, is also equivalent to the Euler-Lagrange
equation for $\lambda $ - this will require some extra care in the choice of
the Lepage equivalent, as we will see later.}.

\item \textit{Symmetries of the Euler-Lagrange form}, \cite{Betounes}, \cite%
{Betounes-symmetry}. Assuming that a fiber preserving diffeomorphism$\
\alpha :Y\rightarrow Y$ is a symmetry of the Euler-Lagrange form $\mathcal{E}%
_{\lambda }=p_{1}d\theta _{\lambda }$, i.e., $J^{s}\alpha ^{\ast }\mathcal{E}%
_{\lambda }=\mathcal{E}_{\lambda }$ for some given Lagrangian $\lambda \in
\Omega _{n,X}(W^{r}),$ then $\lambda $ and $\lambda ^{\prime }:=J^{r}\alpha
^{\ast }\lambda $ must differ by a trivial Lagrangian, which means, by (\ref%
{equality_d_theta}) that: 
\begin{equation}
d\theta _{J^{r}\alpha ^{\ast }\lambda }=d\theta _{\lambda }.
\label{symmetry_EL_form_consequence}
\end{equation}%
If, in addition, $\theta $ obeys the so-called \textit{mapping property}: $%
\theta _{J^{r}\alpha ^{\ast }\lambda }=J^{r}\alpha ^{\ast }\theta _{\lambda
},$ then, one obtains an even stronger statement:\ any symmetry of $\lambda $
is a symmetry of $\theta _{\lambda }$ and any symmetry of $\mathcal{E}%
_{\lambda }$ will also be a symmetry of $d\theta _{\lambda }$ (the converse
implications are true for \textit{any} Lepage equivalent $\theta _{\lambda }$%
).
\end{enumerate}

\bigskip

\textbf{Example:\ The fundamental (Krupka) Lepage equivalent for first order
Lagrangians. }For $\lambda \in \Omega _{n,X}(J^{1}Y),$ a globally defined,
first order Lepage equivalent possessing the closure (and also, the mapping)
property is, \cite{Rossi}, \cite{Saunders}:\textit{\ }%
\begin{equation}
\rho _{\lambda }=\mathcal{L}\omega _{0}+\underset{k=1}{\overset{\min \left\{
m,n\right\} }{\sum }}\dfrac{1}{\left( k!\right) ^{2}}\dfrac{\partial ^{k}%
\mathcal{L}}{\partial y_{~i_{1}}^{\sigma _{1}}...\partial y_{~i_{k}}^{\sigma
_{k}}}\omega ^{\sigma _{1}}\wedge ...\wedge \omega ^{\sigma _{k}}\wedge
\omega _{i_{1}...i_{k}};  \label{fundamental_Lepage_equiv}
\end{equation}%
the degree of contactness of $\rho _{\lambda }$ is $\min \left\{ m,n\right\}
.$ In the case of second order Lagrangians, an extension of the fundamental
Lepage equivalent was recently proposed in \cite{Palese}, but it is not
known whether it has the closure property or not.

\subsection{The Vainberg-Tonti Lagrangian of a source form}

Given a source form $\varepsilon $ defined on some fibered chart domain $%
V^{r}\subset J^{r}Y$, one can canonically attach to $\varepsilon $ and to
the respective chart a Lagrangian called the Vainberg-Tonti Lagrangian, with
the following property:\ if the source form $\varepsilon $ admits a
Lagrangian on $V^{r}$, then the Vainberg-Tonti Lagrangian is a Lagrangian
for $\varepsilon $. The results in this subsection can be found in more
detail in Sections 4.9 and 2.7 of \cite{Krupka-book}.

Consider a fibered chart $(V,\psi )$ on $Y$ and denote $V^{r}:=J^{r}V$.%
\textit{\ }We assume that the image $\psi (V)\subset \mathbb{R}^{m+n}$ is 
\textit{vertically star-shaped}, i.e., for any $(x^{i},y^{\mu })\in \psi
(V), $ the whole segment $\left( x^{i},ty^{\mu }\right) $, $t\in \lbrack
0,1] $ remains in $\psi (V).$ Under this assumption\footnote{%
The construction can also be extended to cases when $\psi (V)$ is not
vertically star-shaped, see, e.g., \cite{Gauss-Bonnet}.}, the correspondence%
\begin{equation}
\chi :(t,(x^{i},y^{\sigma },y_{~i}^{\sigma },...,y_{~i_{1}...i_{r}}^{\sigma
}))\mapsto (x^{i},ty^{\sigma },ty_{~i}^{\sigma
},...,ty_{~i_{1}...i_{r}}^{\sigma })  \label{chi}
\end{equation}%
gives rise to a well defined mapping $\chi :\left[ 0,1\right] \times
V^{r}\rightarrow V^{r}.$ Further, for any $\rho \in \Omega _{k}(V^{r}),$ set:%
\begin{equation}
I\rho :=\overset{1}{\underset{0}{\int }}\rho ^{\left( 0\right) }(t)dt,
\label{I_operator}
\end{equation}%
where $\rho ^{\left( 0\right) }(t)\in \Omega _{k-1}(V^{r})$ is obtained from
the decomposition:%
\begin{equation}
\chi ^{\ast }\rho =dt\wedge \rho ^{\left( 0\right) }(t)+\rho ^{\prime }(t)
\label{hi}
\end{equation}%
into a $dt$-term and a term $\rho ^{\prime }(t)$ which does not contain $dt.$
The obtained mapping $I:\Omega _{k}(V^{r})\rightarrow \Omega _{k-1}(V^{r}),$
called the \textit{fibered homotopy operator}, is $\mathbb{R}$-linear and
obeys:%
\begin{equation}
\rho =Id\rho +dI\rho +\left( \pi ^{r}\right) ^{\ast }\rho _{0},
\label{homotopy_rel}
\end{equation}%
where%
\begin{equation}
\rho _{0}:=0^{\ast }\rho  \label{rho_0}
\end{equation}%
and $0$ denotes the zero section $0:\left( x^{i}\right) \mapsto \left(
x^{i},0,0,...,0\right) $ of $V^{r}.$ The $k$-form $\rho _{0}$ is defined
over $\pi (V)\subset X.$

The following properties will be useful in the following:%
\begin{equation}
Ih\rho =0,~\ \ \ Ip_{k}\rho =p_{k-1}I\rho ,~\ \ \ \ \ 1\leq k\leq q.
\label{homotopy_prop_2}
\end{equation}

\bigskip

Applying the above operator $I\ $to a source form $\varepsilon =\varepsilon
_{\sigma }\omega ^{\sigma }\wedge \omega _{0}\in \Omega _{n+1}(V^{r}),$ the
obtained $n$-form%
\begin{equation}
\lambda _{\varepsilon }:=I\varepsilon  \label{VT_Lagrangian_homotopy}
\end{equation}%
is a Lagrangian on $V^{r},$ called the \textit{Vainberg-Tonti Lagrangian}
attached to $\varepsilon .$ In coordinates: 
\begin{equation}
\lambda _{\varepsilon }=\mathcal{L}_{0}\omega _{0},\text{ \ \ \ \ \ }%
\mathcal{L}_{0}=y^{\sigma }\overset{1}{\underset{0}{\int }}\varepsilon
_{\sigma }(x^{i},ty^{\mu },ty_{~i}^{\mu },...ty_{~i_{1}...i_{r}}^{\mu })dt.
\label{associated_VT_Lagrangian}
\end{equation}%
If the source form $\varepsilon $ admits a Lagrangian on $V$, then: $%
\mathcal{E}_{\lambda _{\varepsilon }}=\varepsilon .$

\bigskip

\textbf{Remark. }The Vainberg-Tonti Lagrangian $\lambda _{\varepsilon }$ of
a source form $\varepsilon $ is of the same order as $\varepsilon ;$ in
particular, for a second order source form, $\lambda _{\varepsilon }$ is
also of second order. That is, very often (e.g., in classical mechanics), $%
\lambda _{\varepsilon }$ can be order-reduced, i.e., it is equivalent to a
lower order Lagrangian.

\section{Canonical Lepage equivalent\label{Section: canonical}}

\subsection{Definition and properties}

In the following, for a given Lagrangian $\lambda $ of arbitrary order $r$,
we will build a local Lepage equivalent $\Phi _{\lambda },$ possessing the
closure property; the obtained Lepage equivalent is 1-contact and of order $%
4r-2.$

\bigskip

Fix a fibered coordinate chart $\left( V,\psi \right) $ as above and
arbitrary Lagrangian $\lambda \in \Omega _{n,X}(V^{r})$ of order $r\geq 2$.
As, by definition, $\lambda $ is a Lagrangian for its own Euler-Lagrange
form $\mathcal{E}_{\lambda }=E_{\sigma }\omega ^{\sigma }\wedge \omega _{0}$%
, the Vainberg-Tonti Lagrangian (of order $\leq 2r$)%
\begin{equation}
\lambda _{VT}:=I\mathcal{E}_{\lambda }  \label{VT_Lagrangian_lambda}
\end{equation}%
is always equivalent to $\lambda .$ The difference between $\lambda $ and $%
\lambda _{VT}$ is thus a trivial Lagrangian; more precisely, one can write,
see Lemma 8, Sec. 4.9 of \cite{Krupka-book}:%
\begin{equation}
\left( \pi ^{2r,r}\right) ^{\ast }\lambda =\lambda _{VT}+hd\alpha ,
\label{lambda_splitting}
\end{equation}%
where%
\begin{equation}
\alpha :=I\Theta _{\lambda }+\left( \pi ^{2r-1}\right) ^{\ast }\mu _{0}
\label{alpha}
\end{equation}%
and $\mu _{0}$ is an $(n-1)$-form on $\pi (V)\subset X$ such that%
\begin{equation}
0^{\ast }\Theta _{\lambda }=d\mu _{0}  \label{mu_0}
\end{equation}%
($\mu _{0}$ is guaranteed to exist, as $0^{\ast }\Theta _{\lambda }$ is a
form of maximal degree on $X$).

\bigskip

We will call the Lagrangian $\lambda _{VT},$ the Vainberg-Tonti Lagrangian%
\textit{\ associated} to $\lambda .$

\bigskip

Let us make the following remarks.

\begin{enumerate}
\item As $\Theta _{\lambda }$ is 1-contact, we obtain by (\ref%
{homotopy_prop_2}) that $\alpha $ is horizontal; moreover, since $\Theta
_{\lambda }$ is generally of order $2r-1$, it follows that $\alpha \in
\Omega _{n,X}(V^{2r-1}),$ i.e., its coordinate expression is: 
\begin{equation*}
\alpha =\alpha ^{i}\omega _{i},~\ \ \ \ \alpha ^{i}=\alpha
^{i}(x^{j},y_{~j}^{\sigma },...,y_{~j_{1}...j_{2r-1}}^{\sigma }).
\end{equation*}

\item For a Lagrangian $\lambda $ of order $r,$ the Euler-Lagrange
expressions (\ref{Euler-Lagrange_expr}) are of order $\leq 2r,$ but their
dependence on the variables $y_{~i_{1}...i_{2r}}^{\sigma }$ is, in any
fibered chart, at most affine. Hence, the associated Vainberg-Tonti
Lagrangian $\lambda _{VT}$ is also at most affine in $y_{~i_{1}...i_{2r}}^{%
\sigma }.$ Consequently, using Proposition \ref{remark_order_Theta}, we find
out that the order of the principal Lepage equivalent $\Theta _{\lambda
_{VT}}$ does not exceed $4r-2.$
\end{enumerate}

We are now able to prove the following result.

\begin{theorem}
\label{canonical_Lepage_equiv}Let $\lambda \in \Omega _{n,X}(V^{r})$ be an
arbitrary Lagrangian of order $r,$ over the vertically star-shaped fibered
chart domain $V$ and $\lambda _{VT}=I\mathcal{E}_{\lambda }\in \Omega
_{n,X}(V^{2r}),$ its associated Vainberg-Tonti Lagrangian. Then:

\begin{enumerate}
\item[\textit{(i)}] The differential form $\Phi _{\lambda }\in \Omega
_{n}(V^{4r-2})$ given by: 
\begin{equation}
\Phi _{\lambda }:=\Theta _{\lambda _{VT}}+\left( \pi ^{4r-2,2r-1}\right)
^{\ast }d\alpha ,  \label{canonical Lepage}
\end{equation}%
where $\alpha $ is given by (\ref{alpha})-(\ref{mu_0}), is a Lepage
equivalent of $\lambda ;$

\item[\textit{(ii)}] If $\lambda $ is a trivial Lagrangian, then $d\Phi
_{\lambda }=0.$
\end{enumerate}
\end{theorem}

\begin{proof}
\textit{(i) }Write $\lambda $ as in (\ref{lambda_splitting}). Then, since
the horizontalization $h$ is a linear mapping, we have, up to the
corresponding jet projections: 
\begin{equation*}
h\Phi _{\lambda }=h\Theta _{\lambda _{VT}}+hd\alpha =\lambda _{VT}+hd\alpha
=\lambda ;
\end{equation*}%
moreover, taking the exterior derivative of (\ref{lambda_splitting}), we
obtain: $d\Phi _{\lambda }=d\Theta _{\lambda _{VT}},$ therefore,%
\begin{equation*}
p_{1}d\Phi _{\lambda }=p_{1}d\Theta _{\lambda _{VT}}=\mathcal{E}_{\lambda
_{VT}}=\mathcal{E}_{\lambda },
\end{equation*}%
which proves that $\Phi _{\lambda }$ is a Lepage equivalent of $\lambda .$

\textit{(ii) }Assuming that $\lambda $ is trivial, we have $\mathcal{E}%
_{\lambda }=0,$ which implies $\lambda _{VT}=0$ and, accordingly, $\Theta
_{\lambda _{VT}}=0;$ as a consequence, $\Phi _{\lambda }=\left( \pi
^{4r-2,2r-1}\right) ^{\ast }d\alpha $ is locally exact - therefore, closed.
\end{proof}

We will call the differential form $\Phi _{\lambda }$ in (\ref%
{lambda_splitting})-(\ref{canonical Lepage}), the \textit{canonical }Lepage
equivalent of $\lambda .$

\bigskip

\textbf{Remarks.}

\begin{enumerate}
\item \textit{(Uniqueness of }$\Phi _{\lambda }$): Though the $\left(
n-1\right) $-form $\mu _{0}$ in (\ref{mu_0}) is not unique, in the
expression of $\Phi _{\lambda },$ it only appears through%
\begin{equation*}
d\alpha =dI\Theta _{\lambda }+\left( \pi ^{2r-1}\right) ^{\ast }d\mu
_{0}=dI\Theta _{\lambda }+\left( \pi ^{2r-1}\right) ^{\ast }0^{\ast }\Theta
_{\lambda }
\end{equation*}
which is uniquely defined.

\item \textit{Linearity of }$\Phi $: All the mappings ($I,\Theta ,\mathcal{E}
$) involved in constructing $\Phi $ are $\mathbb{R}$-linear ones, therefore,%
\begin{equation*}
\Phi :\Omega _{n,X}(V^{r})\mapsto \Omega _{n}(V^{4r-2}),~\ \ \lambda \mapsto
\Phi _{\lambda }
\end{equation*}%
is also an $\mathbb{R}$-linear mapping. Together with the closure property,
this ensures that, for equivalent Lagrangians $\lambda _{1},\lambda _{2},$
we will have $d\Phi _{\lambda _{1}}=d\Phi _{\lambda _{2}}.$
\end{enumerate}

\bigskip

\textbf{Lagrangians admitting globally defined canonical Lepage equivalents. 
}As the above construction heavily relies on quantities that are defined on
a specified chart, such as the Vainberg-Tonti Lagrangian and the principal
Lepage equivalent, a natural question is whether (and when) could $\Phi
_{\lambda }$ be globally defined. Though a complete answer to this question
is out of the scope of this paper, here is a result which covers a lot of
the situations of interest for physical applications.

\begin{theorem}
\label{Theorem_globally_defined}Assume that $Y$ is a tensor bundle over $X$
and $\lambda \in \Omega _{n,X}(J^{r}Y)$ is a Lagrangian of order at most 2,
having second order Lagrange equations. Then, the canonical Lepage
equivalent (\ref{canonical Lepage}) is globally well defined.
\end{theorem}

\begin{proof}
Let us start by the following remark on the fibered homotopy operator $I$
introduced in (\ref{I_operator}). In the particular case when $(Y,\pi ,X)$
has a vector bundle structure, the fiber rescalings $\chi _{t}=\chi \left(
t,\cdot \right) ,$ $t\in \mathbb{R},$ are nothing but the jet prolongations
of the fiberwise scalar multiplication $v\mapsto tv$ on $Y,$ i.e., they make
sense globally on $J^{r}Y$. Accordingly, $\chi :\mathbb{R}\times
J^{r}Y\rightarrow J^{r}Y,$ $\left( t,J_{x}^{r}\gamma \right) \mapsto \chi
_{t}(J_{x}^{r}\gamma )$ is a well defined, smooth mapping. Hence, for any
globally defined form $\rho \in \Omega (J^{r}Y)$, $\chi ^{\ast }\rho $ is
also globally defined on $\mathbb{R}\times J^{r}Y.$ Further, noticing that,
in (\ref{hi}), we can actually write $\rho ^{\left( 0\right) }=\mathbf{i}%
_{\partial _{t}}(\chi ^{\ast }\rho ),$ we obtain%
\begin{equation*}
I\rho =\overset{1}{\underset{0}{\int }}\mathbf{i}_{\partial _{t}}(\chi
^{\ast }\rho )dt,
\end{equation*}%
where, in this case, all the involved operations make sense globally.
Therefore, on vector bundles, $I\rho $ is globally defined.

Assume now that $\lambda $ satisfies the above hypotheses; as $\lambda $ is
globally defined, its Euler-Lagrange form $\mathcal{E}_{\lambda }$ is also
globally defined. Using the above remark, we get that $\lambda _{VT}=I%
\mathcal{E}_{\lambda }$ is globally defined - and, according to our
hypothesis, of second order. But, for second order Lagrangians, the
principal Lepage equivalent is globally defined, which means that so is $%
\Theta _{\lambda _{VT}}$.

On the other hand, as the order of $\lambda $ does not exceed two, $\Theta
_{\lambda }$ is globally well defined. Applying again the above remark on
the operator $I$, we finally get that $d\alpha =dI\Theta _{\lambda }+\left(
\pi ^{2r-1}\right) ^{\ast }0^{\ast }\Theta _{\lambda },$ is also globally
defined. Summing up, we obtain that both terms of $\Phi _{\lambda }$ are
globally defined, which completes the proof.
\end{proof}

\bigskip

The above result applies, for instance, to:

- all generally covariant, \textit{first order} Lagrangians on tensor
bundles;

- Lovelock gravity, Horndeski theories, metric-affine gravity theories with
second order field equations.

\bigskip

\textbf{Example. }Two interesting concrete examples of Lagrangians
satisfying the hypotheses of Theorem \ref{Theorem_globally_defined} are:

-\ the Hilbert Lagrangian of general relativity, see \cite%
{canonical-var-completion}, \cite{Gauss-Bonnet};

- the Lagrangian of Gauss-Bonnet gravity, see \cite{Gauss-Bonnet}.

In both these cases, it was proven in the cited papers that: $\lambda
=\lambda _{VT},$ therefore, we will have: 
\begin{equation*}
\Phi _{\lambda }=\Theta _{\lambda }.
\end{equation*}

\bigskip

Yet, for general Lagrangians, the principal and the canonical Lepage
equivalents will not coincide. Actually, the difference $\Phi _{\lambda
}-\Theta _{\lambda }$ measures the failure of $\Theta _{\lambda }$ from
having the closure property, as shown below.

\begin{proposition}
For a Lagrangian $\lambda =\lambda _{VT}+hd\alpha $ as in (\ref%
{lambda_splitting}), there holds, up to the corresponding jet projections:%
\begin{equation}
\Phi _{\lambda }=\Theta _{\lambda }+(d\alpha -\Theta _{hd\alpha }).
\label{rel_Theta_Phi}
\end{equation}
\end{proposition}

\begin{proof}
From the linearity of $\Theta ,$ we have:\ $\Theta _{\lambda }=\Theta
_{\lambda _{VT}}+\Theta _{hd\alpha }.$ Adding to both hand sides $d\alpha $
and taking into account that, up to jet projections, $\Phi _{\lambda
}=\Theta _{\lambda _{VT}}+d\alpha $, this leads to (\ref{rel_Theta_Phi}).
\end{proof}

\bigskip

The term $d\alpha -\Theta _{hd\alpha }$ in (\ref{rel_Theta_Phi}) is
1-contact. Therefore, using (\ref{1-contact_Lepage_equivs}), there exists a
1-contact form $\nu $ such that, up to the corresponding jet projections: 
\begin{equation}
d\alpha -\Theta _{hd\alpha }=p_{1}d\nu .  \label{nu_alpha}
\end{equation}

\subsection{First order Lagrangians}

For first order\textit{\ }Lagrangians, the canonical Lepage equivalent will
be of order $4r-2=2$. In the following, for this particular case, we will
determine the precise coordinate expression of $\nu $ in (\ref{nu_alpha}).

\begin{proposition}
For an arbitrary first order Lagrangian $\lambda \in \Omega _{n,X}(V^{1}),$
the following statements hold:

\textit{(i) }$\lambda =\lambda _{VT}+\left( d_{i}\alpha ^{i}\right) \omega
_{0},$ where $\alpha ^{i}=\alpha ^{i}(x^{j},y^{\sigma },y_{~j}^{\sigma })$
are of order 1;

(ii) the canonical and the principal Lepage equivalents of $\lambda $ are
related by:%
\begin{equation}
\Phi _{\lambda }=\left( \pi ^{2,1}\right) ^{\ast }\Theta _{\lambda
}+p_{1}d\nu ,  \label{first_order_rel_Theta_Phi}
\end{equation}%
where%
\begin{equation}
\nu :=\dfrac{1}{4}(\dfrac{\partial \alpha ^{j}}{\partial y_{~i}^{\sigma }}-%
\dfrac{\partial \alpha ^{i}}{\partial y_{~j}^{\sigma }})\omega ^{\sigma
}\wedge \omega _{ij}.  \label{nu}
\end{equation}
\end{proposition}

\begin{proof}
\textit{(i) }The form $\alpha =I\Theta _{\lambda }+\mu _{0}$ is of order 1
and $\pi ^{1}$-horizontal, therefore it is locally expressed as $\alpha
=\alpha ^{i}\omega _{i},$ with $\alpha ^{i}=\alpha ^{i}(x^{j},y^{\sigma
},y_{~j}^{\sigma })$ only (the precise expression of $\alpha ^{i}$ can be
found by applying (\ref{I_operator}) to $\Theta _{\lambda },$ but it is less
essential in the following). The statement then follows from $hd\alpha
=\left( d_{i}\alpha ^{i}\right) \omega _{0}.$

\textit{(ii) }With $\alpha $ as above, we have: $\Phi _{\lambda }=\Theta
_{\lambda _{VT}}+\left( \pi ^{2,1}\right) ^{\ast }d\alpha .$ The Lagrangian $%
hd\alpha =:\mathcal{L}_{0}\omega _{0}$ (where $\mathcal{L}_{0}=d_{k}\alpha
^{k}$) is of second order, therefore its principal Lepage equivalent is
expressed as%
\begin{equation*}
\Theta _{hd\alpha }=\mathcal{L}_{0}\omega _{0}+B_{~\sigma }^{i}\omega
^{\sigma }\wedge \omega _{i}+B_{~\sigma }^{ij}\omega _{~j}^{\sigma }\wedge
\omega _{i},
\end{equation*}%
where $B_{~\sigma }^{i},B_{~\sigma }^{ij}$ are given by (\ref{B_general}).
That is, 
\begin{equation}
B_{~\sigma }^{ij}=\dfrac{\partial \mathcal{L}_{0}}{\partial y_{~ij}^{\sigma }%
}=\dfrac{\partial }{\partial y_{~ij}^{\sigma }}(d_{k}\alpha ^{k})=\dfrac{1}{2%
}(\dfrac{\partial \alpha ^{i}}{\partial y_{~j}^{\sigma }}+\dfrac{\partial
\alpha ^{j}}{\partial y_{~i}^{\sigma }})  \label{B2}
\end{equation}%
and from $\dfrac{\partial \mathcal{L}_{0}}{\partial y_{~i}^{\sigma }}=\dfrac{%
\partial }{\partial y_{~i}^{\sigma }}(d_{k}\alpha ^{k})=d_{k}(\dfrac{%
\partial \alpha ^{k}}{\partial y_{~i}^{\sigma }})+\dfrac{\partial \alpha ^{i}%
}{\partial y^{\sigma }}$, we find:%
\begin{equation*}
B_{~\sigma }^{i}=\dfrac{\partial \alpha ^{i}}{\partial y^{\sigma }}+\dfrac{1%
}{2}d_{j}(\dfrac{\partial \alpha ^{j}}{\partial y_{~i}^{\sigma }}-\dfrac{%
\partial \alpha ^{i}}{\partial y_{~j}^{\sigma }}).
\end{equation*}%
Substituting into $\Theta _{hd\alpha }$ and taking into account that $\left(
\pi ^{2,1}\right) ^{\ast }d\alpha =(d_{k}\alpha ^{k})\omega _{0}+\dfrac{%
\partial \alpha ^{i}}{\partial y^{\sigma }}\omega ^{\sigma }\wedge \omega
_{i}+\dfrac{\partial \alpha ^{i}}{\partial y_{~j}^{\sigma }}\omega
_{~j}^{\sigma }\wedge \omega _{i},$ we finally get:%
\begin{equation}
\left( \pi ^{2,1}\right) ^{\ast }d\alpha -\Theta _{hd\alpha }=\dfrac{1}{2}%
d_{j}(\dfrac{\partial \alpha ^{i}}{\partial y_{~j}^{\sigma }}-\dfrac{%
\partial \alpha ^{j}}{\partial y_{~i}^{\sigma }})\omega ^{\sigma }\wedge
\omega _{i}+\dfrac{1}{2}(\dfrac{\partial \alpha ^{i}}{\partial
y_{~j}^{\sigma }}-\dfrac{\partial \alpha ^{j}}{\partial y_{~i}^{\sigma }}%
)\omega _{~j}^{\sigma }\wedge \omega _{i}.  \label{aux}
\end{equation}%
On the other hand, a direct computation using: $d\omega ^{\sigma }=-\omega
_{~k}^{\sigma }\wedge dx^{k}$ and $dx^{k}\wedge \omega _{ij}=\delta
_{~j}^{k}\omega _{i}-\delta _{i}^{k}\omega _{j}$ shows that $p_{1}d\nu $ is
precisely the right hand side of the above. The statement then follows from (%
\ref{rel_Theta_Phi}).
\end{proof}

\section{Reduced Lepage equivalents\label{Section: minimal}}

In the following, we present an alternative construction, which is
advantageous in the case when the Lagrangian $\lambda $ can be
order-reduced; for reducibility criteria, see, e.g., \cite{Grigore} \cite%
{Rossi}, \cite{Saunders-reducibility}.

Consider a Lagrangian $\lambda \in \Omega _{n,X}(W^{r}),$ where $%
W^{r}\subset J^{r}Y$ is open, and pick any equivalent Lagrangian $\lambda
^{\prime }\in \Omega _{n,X}(W^{s})$ to $\lambda ,$ of \textit{minimal} order 
$s\leq r.$ Then, again, we can write%
\begin{equation}
\lambda =(\pi ^{r,s})^{\ast }\lambda ^{\prime }+hd\alpha ,
\label{minimal_order_decomp}
\end{equation}%
for some $\alpha \in \Omega _{n-1}(W^{r-1})$.

In particular, for a trivial Lagrangian $\lambda ,$ minimal order
Lagrangians equivalent to $\lambda $ are $\pi ^{r}$-projectable $n$-forms $%
\lambda ^{\prime }=f(x^{i})\omega _{0}$.

\begin{proposition}
\label{Th_minimal_Lepage_equiv}Let $\lambda \in \Omega _{n,X}(W^{r})$ be an
arbitrary Lagrangian and $\lambda ^{\prime }\in \Omega _{n,X}(W^{s}),$ a
dynamically equivalent Lagrangian to $\lambda ,$ of minimal order $s\leq r.$
Then:

(i) The $n$-form 
\begin{equation}
\phi _{\lambda }:=\Theta _{\lambda ^{\prime }}+d\alpha ,
\label{reduced_Lepage_equiv}
\end{equation}%
where $\lambda ^{\prime }$ and $\alpha $ are as in (\ref%
{minimal_order_decomp}) and the equality must be understood up to the
corresponding jet projections, is a Lepage equivalent of $\lambda .$

(ii) If $\lambda $ is variationally trivial, then any $\phi _{\lambda }$
constructed as above is closed.
\end{proposition}

\begin{proof}
\textit{(i) }The proof is similar to the one of Theorem \ref%
{canonical_Lepage_equiv}. First, we note that, up to jet projections:%
\begin{equation*}
h\phi _{\lambda }=h\Theta _{\lambda ^{\prime }}+hd\alpha =\lambda ^{\prime
}+hd\alpha =\lambda ;
\end{equation*}%
moreover, $d\phi _{\lambda }=d\Theta _{\lambda ^{\prime }}$ implies $%
p_{1}d\phi _{\lambda }=p_{1}d\Theta _{\lambda ^{\prime }}=\mathcal{E}%
_{\lambda ^{\prime }}=\mathcal{E}_{\lambda },$ which is a source form, that
is, $\phi _{\lambda }$ is a Lepage equivalent of $\lambda .$

\textit{(ii) }If $\lambda $ is trivial, then $\lambda ^{\prime
}=f(x^{i})\omega _{0}$, which gives: $\Theta _{\lambda ^{\prime }}=\lambda
^{\prime }.$ But, as $\lambda ^{\prime }$ is an $n$-form on $X,$ $\dim X=n,$
we find that: $d\phi _{\lambda }=d\Theta _{\lambda ^{\prime }}=d\lambda
^{\prime }=0.$
\end{proof}

\bigskip

\begin{definition}
We will call any Lepage equivalent built as in (\ref{minimal_order_decomp})-(%
\ref{reduced_Lepage_equiv}), a reduced Lepage equivalent of $\lambda .$
\end{definition}

\textbf{Remarks.}

\begin{enumerate}
\item The reduced Lagrangian $\lambda ^{\prime }$ of $\lambda $ (if it
exists) is, generally, not unique. As a consequence, we may obtain multiple
reduced Lepage equivalents $\phi _{\lambda }$ for the same Lagrangian. Even
so, the multi-valued correspondence $\lambda \mapsto \phi _{\lambda }$ is $%
\mathbb{R}$-linear, in the following sense: for any $\lambda _{1},\lambda
_{2}\in \Omega _{n,X}(W^{r})$ and $a_{1},a_{2}\in \mathbb{R}$, if $\phi
_{\lambda _{1}}\in \phi \left( \lambda _{1}\right) $ and $\phi _{\lambda
_{2}}\in \phi \left( \lambda _{2}\right) ,$ then $a_{1}\phi _{\lambda
_{1}}+a_{2}\phi _{\lambda _{2}}$ belongs to the image $\phi \left(
a_{1}\lambda _{1}+a_{2}\lambda _{2}\right) $.

\item The splitting (\ref{minimal_order_decomp}) is, generally, only local -
therefore, reduced Lepage equivalents are, in general, defined only locally.
\end{enumerate}

\bigskip

In particular, for second order Lagrangians, we obtain:

\begin{proposition}
Any reducible second order Lagrangian admits a local first order Lepage
equivalent.
\end{proposition}

\begin{proof}
If $\lambda \in \Omega _{n,X}(W^{2})$ is reducible to a first order
Lagrangian $\lambda ^{\prime }\in \Omega _{n,X}(W^{1}),$ the corresponding
reduced Lepage equivalent $\phi _{\lambda }$ is of order 1, as both $\Theta
_{\lambda ^{\prime }}$ and $\alpha $ are, in this case, of order 1.
\end{proof}

\section{Examples}

\subsection{Hilbert Lagrangian}

The Hilbert Lagrangian $\lambda _{g}$\textit{\ }is a peculiar example; in
this case, we will see below that: 
\begin{equation}
\Phi _{\lambda _{g}}=\Theta _{\lambda _{g}}=\phi _{\lambda _{g}},
\label{Hilbert}
\end{equation}%
where $\phi _{\lambda _{g}}$ corresponds to the famous non-invariant, first
order Lagrangian equivalent to $\lambda _{g}$. In particular, $\Phi
_{\lambda _{g}}$ is of order 1.

The first equality above has already been discussed in Section \ref{Section:
canonical}. The second one is based on a result in \cite%
{inverse-problem-book}, as follows. Denote by $Y=Met(X)$, the bundle of
nondegenerate tensors of type (0,2)\ over $X$ and by $%
(g_{ij};g_{ij,k};g_{ij,kl}),$ the coordinates in a fibered chart on $J^{2}Y.$
The Hilbert Lagrangian%
\begin{equation*}
\lambda _{g}=\mathcal{R}\omega _{0},~\ ~\ \mathcal{R}:=R\sqrt{\left\vert
\det g\right\vert }
\end{equation*}%
can be split as $\lambda _{g}=\lambda _{g}^{\prime }+\left( d_{i}\alpha
^{i}\right) \omega _{0},$ where%
\begin{equation}
\lambda _{g}^{\prime }=g^{jk}(\Gamma _{~jl}^{i}\Gamma _{~ki}^{l}-\Gamma
_{~jk}^{i}\Gamma _{~il}^{l})\sqrt{\left\vert \det g\right\vert }\omega _{0}=:%
\mathcal{L}_{g}^{\prime }\omega _{0}  \label{reduced_Hilbert_Lagrangian}
\end{equation}%
is the reduced (non-invariant) Lagrangian for $\lambda _{g}$ and: 
\begin{equation*}
\alpha =(\Gamma _{~j}^{ij}-\Gamma _{~j}^{ji})\sqrt{\left\vert \det
g\right\vert }\omega _{i}.
\end{equation*}%
In \textbf{\cite{inverse-problem-book},} Sec. 5.5.1, it was shown that:%
\begin{equation}
\Theta _{\lambda _{g}}=\Theta _{\lambda _{g}^{\prime }}+d\alpha ,
\label{Theta_Hilbert}
\end{equation}%
which proves the second equality (\ref{Hilbert}).

The coordinate expression of $\Theta _{\lambda _{g}}$ in the natural basis $%
\left\{ dx^{i},dg_{jk},dg_{jk,i}\right\} $ of $\Omega (J^{1}Y)$ is known, 
\cite{KKS}, as:%
\begin{eqnarray*}
\Theta _{\lambda _{g}} &=&g^{ip}(\Gamma _{~ip}^{j}\Gamma _{~jk}^{k}-\Gamma
_{~ik}^{j}\Gamma _{~jp}^{k})\sqrt{\left\vert \det g\right\vert }\omega _{0}
\\
&&+\left( g^{jp}g^{iq}-g^{pq}g^{ij}\right) \sqrt{\left\vert \det
g\right\vert }(dg_{pq,j}+\Gamma _{~pq}^{k}dg_{jk})\wedge \omega _{i}.
\end{eqnarray*}%
Let us explicitly calculate in the following the term $\Theta _{\lambda
_{g}^{\prime }}$ in (\ref{Theta_Hilbert}). This will only contain a $dg_{pq}$%
-component - or, in the contact basis, an $\omega _{\left( pq\right) }$%
-component, where $\omega _{(pq)}=dg_{pq}-g_{pq,r}dx^{r}.$ More precisely, 
\begin{equation*}
\Theta _{\lambda _{g}^{\prime }}=\mathcal{L}_{g}^{\prime }\omega _{0}+\dfrac{%
\partial \mathcal{L}_{g}^{\prime }}{\partial g_{pq,r}}\omega _{(pq)}\wedge
\omega _{r}.
\end{equation*}%
The derivative $\dfrac{\partial \mathcal{L}_{g}^{\prime }}{\partial g_{pq,r}}
$ can be calculated directly, using $\mathcal{L}_{g}^{\prime }=g^{ih}\left(
g^{jk}g^{lm}-g^{jl}g^{mk}\right) \Gamma _{hjl}\Gamma _{mki}$ and $\dfrac{%
\partial \Gamma _{hjl}}{\partial g_{pq,r}}=\dfrac{1}{2}(\delta
_{h}^{p}\delta _{j}^{q}\delta _{l}^{r}+\delta _{h}^{p}\delta _{l}^{q}\delta
_{j}^{r}-\delta _{j}^{p}\delta _{l}^{q}\delta _{h}^{r}).$ We obtain:%
\begin{equation*}
\dfrac{\partial \mathcal{L}_{g}^{\prime }}{\partial g_{pq,r}}=\Gamma
^{rpq}-g^{qr}\Gamma _{~\ k}^{kp}+\dfrac{1}{2}g^{pq}(\Gamma
_{~~j}^{jr}-\Gamma _{~~j}^{rj}).
\end{equation*}

\subsection{Klein-Gordon field Lagrangian}

A simplest and somwehat suprising example is provided by the classical
Klein-Gordon field Lagrangian.

Consider the trivial bundle $Y=\mathbb{R}^{4}\times \mathbb{R},\ $where $X=%
\mathbb{R}^{4}$ is equipped with the Minkowski metric $\eta
=diag(1,-1,-1,-1) $ (and corresponding Cartesian coordinates $\left(
x^{i}\right) _{i=\overline{1,4}}$). We denote a set of global fibered
coordinates on $Y$ by $\left( x^{i},\varphi \right) $ and by $(x^{i},\varphi
,\varphi _{i},\varphi _{ij})$ the induced coordinates on $J^{2}Y.$ The
contact basis elements on $\Omega (J^{2}Y)$ will then be denoted as: $\tilde{%
\omega}=d\varphi -\varphi _{i}dx^{i},$ $\tilde{\omega}_{i}=d\varphi
_{i}-\varphi _{ij}dx^{j}.$

The Klein-Gordon field Lagrangian $\lambda =\mathcal{L}\omega _{0}\in \Omega
_{4}(J^{1}Y)$ is given (see, e.g., \cite{Franklin}) by:%
\begin{equation}
\mathcal{L=}\dfrac{1}{2}(\eta ^{ij}\varphi _{i}\varphi _{j}-m^{2}\varphi
^{2}),  \label{KG Lagrangian}
\end{equation}%
where $m\geq 0$ is a constant.

\begin{itemize}
\item Since the above Lagrangian is of first order (i.e., it is of minimal
order in its equivalence class), its principal Lepage equivalent is also a
minimal one. We immediately obtain:%
\begin{equation*}
\phi _{\lambda }=\Theta _{\lambda }=\mathcal{L}\omega _{0}+(\eta
^{ij}\varphi _{j})\tilde{\omega}\wedge \omega _{i}.
\end{equation*}

\item As $\lambda $ is of order 1, the canonical Lepage equivalent $\Phi
_{\lambda }$ can be most easily determined from (\ref%
{first_order_rel_Theta_Phi})-(\ref{nu}).

The unique Euler-Lagrange expression $\mathcal{E}:=\mathcal{E}_{\sigma }$ of 
$\lambda $ is: $\mathcal{E}=-(\eta ^{ij}\varphi _{ij}+m^{2}\varphi ),$ which
allows us to calculate the associated Vainberg-Tonti Lagrangian $\lambda
_{VT}=\mathcal{L}_{VT}\omega _{0}$ of $\lambda $ as: 
\begin{equation*}
\mathcal{L}_{VT}=-\varphi \overset{1}{\underset{0}{\int }}(\eta
^{ij}t\varphi _{ij}+m^{2}t\varphi )dt=-\dfrac{1}{2}(\eta ^{ij}\varphi
\varphi _{ij}+m^{2}\varphi ^{2});
\end{equation*}%
this gives: $\mathcal{L=L}_{VT}+(d_{i}\alpha ^{i}),$ with:%
\begin{equation*}
\alpha ^{i}=\dfrac{1}{2}\eta ^{ij}\varphi \varphi _{j}.
\end{equation*}%
Noticing that $\dfrac{\partial \alpha ^{i}}{\partial \varphi _{j}}=\dfrac{%
\partial \alpha ^{j}}{\partial \varphi _{i}}=\dfrac{1}{2}\eta ^{ij}\varphi ,$
we find $\nu =\dfrac{1}{4}(\dfrac{\partial \alpha ^{j}}{\partial \varphi _{i}%
}-\dfrac{\partial \alpha ^{i}}{\partial \varphi _{j}})\tilde{\omega}\wedge
\omega _{ij}=0$, which means that $\Phi _{\lambda }=\Theta _{\lambda }.$ All
in all, we obtain,%
\begin{equation*}
\Phi _{\lambda }=\Theta _{\lambda }=\varphi _{\lambda },
\end{equation*}%
though, in this case, $\lambda _{VT}$ and $\lambda $ do not coincide.
\end{itemize}

\subsection{Electromagnetic field Lagrangian}

Another standard example of a first order field Lagrangian is the
electromagnetic field Lagrangian (e.g., \cite{Franklin}). Consider the
cotangent bundle $Y=T^{\ast }\mathbb{R}^{4}$ over the Minkowski spacetime $%
\left( \mathbb{R}^{4},\eta \right) .$ We denote the coordinates in a fibered
chart on $J^{2}Y$ by $\left( x^{i};A_{i};A_{i,j};A_{i,jk}\right) $ and the
contact basis 1-forms by $\tilde{\omega}_{i}=dA_{i}-A_{i,j}dx^{j},$ $\tilde{%
\omega}_{i,j}=dA_{i,j}-A_{i,jk}dx^{k}.$

We will investigate the Lagrangian $\lambda =\mathcal{L}\omega _{0}\in
~\Omega _{4,X}(J^{1}Y)$, with:%
\begin{equation*}
\mathcal{L}=F_{ij}F^{ij},
\end{equation*}%
where $F_{ij}=A_{j,i}-A_{i,j}$ and $F^{ij}=\eta ^{ik}\eta ^{jl}F_{kl}$.

\begin{itemize}
\item Noting that $\dfrac{\partial F_{ij}}{\partial A_{k,l}}=\delta
_{j}^{k}\delta _{i}^{l}-\delta _{i}^{k}\delta _{j}^{l}$, we find: 
\begin{equation*}
\dfrac{\partial \mathcal{L}}{\partial A_{j,i}}=4F^{ij},
\end{equation*}%
which gives the principal Lepage equivalent:%
\begin{equation*}
\Theta _{\lambda }=\mathcal{L}\omega _{0}+\dfrac{\partial \mathcal{L}}{%
\partial A_{j,i}}\tilde{\omega}_{j}\wedge \omega _{i}=\mathcal{L}\omega
_{0}+4F^{ij}\tilde{\omega}_{j}\wedge \omega _{i}.
\end{equation*}%
As the Lagrangian is a first order one, this is also a minimal Lepage
equivalent.

A nice feature of $\Theta _{\lambda }$ is that, under gauge transformations $%
A_{i}\mapsto A_{i}+\partial _{i}f$ (with $f=f(x^{k})$ defined on the base
manifold $\mathbb{R}^{4}$), $\Theta _{\lambda }$ remains invariant. This
follows as, on the one hand, the basis contact forms $\tilde{\omega}_{j}$
are easily seen to be invariant (and the same is true for $\tilde{\omega}%
_{j,k},$ though this is unessential for the moment) and on the other hand,
the coefficients $F_{ij}$ are gauge invariant.

\item The Euler-Lagrange expressions%
\begin{equation*}
\mathcal{E}^{j}=\dfrac{\partial \mathcal{L}}{\partial A_{j}}-d_{i}(\dfrac{%
\partial \mathcal{L}}{\partial A_{j,i}})=-4d_{i}F^{ij},\ 
\end{equation*}%
lead to the associated Vainberg-Tonti Lagrangian $\lambda _{VT}=\mathcal{L}%
_{VT}\omega _{0},$ as follows:%
\begin{equation*}
\mathcal{L}_{VT}=-4A_{j}\overset{1}{\underset{0}{\int }}%
d_{i}F^{ij}tdt=-2A_{j}d_{i}F^{ij}.
\end{equation*}%
(where we took into account that $A^{i,j}\circ \chi _{t}=tA^{i,j}$). We thus
obtain the desired divergence expression $d_{i}\alpha ^{i}=\mathcal{L}-%
\mathcal{L}_{VT}$ as:%
\begin{equation*}
d_{i}\alpha
^{i}=F_{ij}F^{ij}+2A_{j}d_{i}F^{ij}=2A_{j,i}F^{ij}+2A_{j}d_{i}F^{ij}=d_{i}(2A_{j}F^{ij}),
\end{equation*}%
which allows us to identify:%
\begin{equation*}
\alpha ^{i}=2A_{l}F^{il}.
\end{equation*}%
We will calculate the canonical Lepage equivalent $\Phi _{\lambda }$ using (%
\ref{first_order_rel_Theta_Phi})-(\ref{nu}); we find that $p_{1}d\nu $ is
given similarly to (\ref{aux}):%
\begin{equation*}
p_{1}d\nu =\dfrac{1}{2}d_{j}(\dfrac{\partial \alpha ^{i}}{\partial A_{k,j}}-%
\dfrac{\partial \alpha ^{j}}{\partial A_{k,i}})\tilde{\omega}_{k}\wedge
\omega _{i}+\dfrac{1}{2}(\dfrac{\partial \alpha ^{i}}{\partial A_{k,j}}-%
\dfrac{\partial \alpha ^{j}}{\partial A_{k,i}})\tilde{\omega}_{k,j}\wedge
\omega _{i}
\end{equation*}%
To this aim, we first note that $\dfrac{\partial A^{l,i}}{\partial A_{k,j}}%
=\eta ^{lk}\eta ^{ij},$ which leads after a brief computation to:%
\begin{equation*}
\dfrac{1}{2}(\dfrac{\partial \alpha ^{i}}{\partial A_{k,j}}-\dfrac{\partial
\alpha ^{j}}{\partial A_{k,i}})=A^{i}\eta ^{jk}-A^{j}\eta ^{ik}
\end{equation*}%
and finally%
\begin{equation*}
p_{1}d\nu =(A^{i,k}-A_{~,j}^{j}\eta ^{ik})\tilde{\omega}_{k}\wedge \omega
_{i}+(A^{i}\eta ^{jk}-A^{j}\eta ^{ik})\tilde{\omega}_{k,j}\wedge \omega _{i}
\end{equation*}%
Thus, the canonical Lepage equivalent $\Phi _{\lambda }=\left( \pi
^{2,1}\right) ^{\ast }\Theta _{\lambda }+p_{1}d\nu $ is of \textit{first}
order:%
\begin{equation*}
\Phi _{\lambda }=\mathcal{L}\omega _{0}+(4F^{ik}+A^{i,k}-A_{~,j}^{j}\eta
^{ik})\tilde{\omega}_{k}\wedge \omega _{i}+\left( A^{i}\eta ^{jk}-A^{j}\eta
^{ik}\right) \tilde{\omega}_{k,j}\wedge \omega _{i}
\end{equation*}%
(as the second order term arising from the presence of $A_{k,jl}$ in $\tilde{%
\omega}_{k,j}=dA_{k,j}-A_{k,jl}dx^{l},$ actually vanishes).

Another interesting remark is the following. $\Phi _{\lambda }$ is \textit{%
not} invariant under gauge transformations $A_{i}\mapsto A_{i}+\partial _{i}f
$, hence, it has less symmetries than the Lagrangian $\lambda $ (in
particular, this points out that $\Phi $ does not possess the mapping
property $J^{r}\alpha ^{\ast }\Phi _{\lambda }=\Phi _{J^{r}\alpha ^{\ast
}\lambda }$). Yet, in the exterior derivative%
\begin{equation*}
d\Phi _{\lambda }=\mathcal{E}_{\lambda }+2(2\eta ^{jk}\eta ^{il}-\eta
^{ij}\eta ^{kl}-\eta ^{ik}\eta ^{jl})\tilde{\omega}_{j,l}\wedge \tilde{\omega%
}_{k}\wedge \omega _{i}
\end{equation*}%
gauge invariance is restored.
\end{itemize}

\section{Conclusion and outlook}

In the present paper, we have proposed two notions of local Lepage
equivalent $\theta _{\lambda }$ for Lagrangians $\lambda $ of arbitrary
order $r\geq 1,$ possessing the so-called \textit{closure property: }the
given Lepage equivalent is a closed differential form if and only if the
Lagrangian $\lambda $ is trivial. Both these notions are constructed as%
\begin{equation}
\theta _{\lambda }=\Theta _{\bar{\lambda}}+d\alpha ,  \label{general_recipe}
\end{equation}%
i.e., by adding an exact form $d\alpha $ to the\textit{\ principal (Poincar%
\'{e}-Cartan) form} $\Theta _{\bar{\lambda}}$ of an appropriately chosen
equivalent Lagrangian $\bar{\lambda}=\lambda -hd\alpha ;$ this guarantees
that \textit{all }equivalent Lagrangians to $\lambda $ will have the same $%
d\theta :$%
\begin{equation}
d\theta _{\lambda }=d\Theta _{\bar{\lambda}}.  \label{d_theta}
\end{equation}%
In the above, we studied the following choices for $\bar{\lambda}$:

\begin{enumerate}
\item The Vainberg-Tonti Lagrangian $\lambda _{VT}=I\mathcal{E}_{\lambda },$
sharing the same Euler-Lagrange form $\mathcal{E}_{\lambda }$ with $\lambda
; $ this Lagrangian is uniquely (and, in a sense, canonically) defined by
the given Lagrangian $\lambda ;$ the obtained Lepage equivalent is
1-contact, yet, generally, of higher order than the principal Lepage
equivalent $\Theta _{\lambda }.$

\item A Lagrangian of minimal order, equivalent to $\lambda $, which leads
to a computationally simplest possible (though, generally not unique) Lepage
equivalent for $\lambda .$ With such a choice, the exterior derivative $%
d\phi _{\lambda }$ takes the simple possible form, i.e., it is always
2-contact, $\omega ^{\sigma }$-generated and of minimal order; such $d\phi
_{\lambda }$ is called in \cite{Rossi}, a \textit{minimal Lepage extension}
of the Euler-Lagrange form $\mathcal{E}_{\lambda }$. In particular, for
reducible second order Lagrangians, our construction gives a recipe for
obtaining a \textit{first order }Lepage equivalent.
\end{enumerate}

In both the above cases (and, actually, for \textit{any }choice of $\lambda $%
), the obtained $d\theta _{\lambda }$ is at most 2-contact - which is the
minimal possible degree of contactness for the exterior derivative of a
Lepage equivalent. Besides simplicity, our algorithm (\ref{general_recipe})
has also the advantage of generality -- as, at least in principle, it might
also allow for other interesting choices of $\bar{\lambda},$ while still
preserving the property (\ref{d_theta}).

Moreover, basing our construction on the principal (Poincar\'{e}-Cartan)
Lepage equivalent as in (\ref{general_recipe})-(\ref{d_theta}) has promising
features, e.g., for Hamiltonian field theory. It is known (see, e.g., \cite%
{Krupkova-Smetanova}), that, under certain regularity conditions on the
Lagrangian function $\mathcal{L},$ the Hamilton equation (\ref%
{coord-free-Hamilton-eq}) for the principal Lepage equivalent $\Theta
_{\lambda }$ (called the \textit{Hamilton-de-Donder }equation) becomes,
indeed, equivalent to the Euler-Lagrange equation of $\lambda ,$ which is
what one would expect from a "correct"\ Hamilton equation. Also, in some
particular cases when the Lagrangian $\lambda $ fails to satisfy the
regularity condition, a "regularization" procedure is proposed in \cite%
{Krupkova-Smetanova} by passing to an equivalent, regular Lagrangian $\bar{%
\lambda}$ and considering the Hamilton equation for $\Theta _{\bar{\lambda}}$
(which is, thus, equivalent to the Euler-Lagrange equation for $\lambda $)
as the relevant Hamilton equation for $\lambda ;$ in this case, choosing a
Lepage equivalent as in (\ref{general_recipe})-(\ref{d_theta}) guarantees
that the obtained Hamilton equation for $\theta _{\lambda }$ is a correct
one. A more detailed study of regularizable Lagrangians and of the resulting
Hamilton theory, using the proposed Lepage equivalents, will be done in the
near future.

\bigskip

\textbf{Acknowledgment. }We are extremely grateful to prof. Demeter Krupka
for drawing our attention towards the closure property topic and for useful
and extensive talks on it and also, to Bence Racsk\'{o} for helpful
discussions on reducible second order Lagrangians.

\end{document}